# SPARSE MATRIX MULTIPLICATION ON AN ASSOCIATIVE PROCESSOR

L. Yavits, A. Morad, R. Ginosar

**Abstract**—Sparse matrix multiplication is an important component of linear algebra computations. Implementing sparse matrix multiplication on an associative processor (AP) enables high level of parallelism, where a row of one matrix is multiplied in parallel with the entire second matrix, and where the execution time of vector dot product does not depend on the vector size. Four sparse matrix multiplication algorithms are explored in this paper, combining AP and baseline CPU processing to various levels. They are evaluated by simulation on a large set of sparse matrices. The computational complexity of sparse matrix multiplication on AP is shown to be an $O(nnz)$ where $nnz$ is the number of nonzero elements. The AP is found to be especially efficient in binary sparse matrix multiplication. AP outperforms conventional solutions in power efficiency.

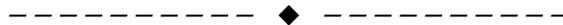

## 1 INTRODUCTION

Sparse matrix multiplication is a frequent bottleneck in large scale linear algebra applications, especially in data mining and machine learning [30]. The efficiency of sparse matrix multiplication becomes even more relevant with the emergence of big data, giving rise to very large vector and matrix sizes.

Associative Processor (AP) is a massively parallel SIMD array processor [15][23][45]. The AP comprises a Content Addressable Memory (CAM) and facilitates processing in addition to storage. The execution time of a typical vector operation in an AP does not depend on the vector size, thus allowing efficient parallel processing of very large vectors. AP's efficiency grows with the data set sizes and data-level parallelism.

Associative processing has been known and extensively studied since the 1960s. Commercial associative processing never quite took off, because only limited amounts of memory could be placed on a single die [22]. Equally important, standalone bit- and word-parallel conventional SIMD processors outperformed APs due to the data sets and tasks of limited size. However, the progress in computer industry and semiconductor technology in recent years opens the door for reconsidering the APs:

- The rise of big data pushes the computational requirements to levels never seen before. The amounts of data to be processed simultaneously require a new parallel computing paradigm.
- Power consumption, which used to be a secondary factor in the past, has become a principal constraint on scalability and performance of the parallel architectures. The AP is shown to achieve a better power efficiency [25].

- On-chip memory bandwidth is another factor limiting the performance and scalability of parallel architectures. Associative processing mitigates this limitation by intertwining computing with data storage.
- In high performance dies, thermal density is becoming the limit on total computation capabilities; associative processing leads to uniform power and thermal distribution over the chip area, avoiding hot spots and enabling the three dimensional (3-D) integration.

In this work, we present four associative algorithms for sparse matrix by dense vector (SpMV) and/or sparse matrix by dense matrix (SpMM) as well as sparse matrix by sparse matrix multiplication. The latter one is less regular than SpMM, making it ever more difficult to map onto a fine-grain massively parallel processor.

The first algorithm, designated "AP", is a fully associative implementation, making use only of the intrinsic AP resources. We show that the computational complexity of a fully associative implementation is $O(nnz)$, where $nnz$ is the number of nonzero elements. In the second algorithm, called "AP+ACC", the singleton products are computed by the AP and an external baseline CPU is used to accumulate them. The third algorithm, "AP+MULT", uses a CPU to multiply matrix elements; the products are accumulated by the AP. The fourth algorithm, "AP+MULT+ACC", uses the AP for matching the matrix elements, and the CPU for both multiplication and accumulation. We find that the fully associative implementation is especially efficient for very large matrices with large number of nonzero elements per row. Fully associative implementation is also preferred for multiplication of binary sparse matrices (that is, matrices where the nonzero elements are $\pm 1$). In contrast, the other three (hybrid) algorithms are more efficient for matrices with a lower number of nonzero elements per row, and their efficiency improves slower or remains constant with the number of nonzero elements.

The rest of this paper is organized as follows. Section 2 discusses the related work. Section 3 presents the archi-

————————————————
- *Leonid Yavits (\*), E-mail: yavits@tx.technion.ac.il.*
- *Amir Morad (\*), E-mail: amirm@tx.technion.ac.il.*
- *Ran Ginosar (\*), E-mail: ran@ee.technion.ac.il.*

*(\*) Authors are with the Department of Electrical Engineering, Technion-Israel Institute of Technology, Haifa 32000, Israel.*

tecture of the associative processor and principles of associative computing. Section 4 presents associative algorithms for sparse matrix multiplication. Section 5 details the evaluation methodology and presents the simulation results. Section 6 offers conclusions.

## 2 RELATED WORK

A majority of previous studies target sparse matrix by dense vector multiplication (SpMV) or sparse matrix by dense matrix multiplication (SpMM). Sparse matrix by sparse matrix multiplication has been rarely addressed in prior research. For simplicity, in this section we apply the term SpMM to both SpMM and SpMV.

A substantial body of literature explores sparse matrix multiplication optimization techniques. A comprehensive review of these techniques is provided by R. Vuduc [38]. We take a slightly different look, focusing on hardware platforms rather than on software implementation. The literature can be divided into three categories, as summarized in TABLE 1.

TABLE 1
RELATED WORK SUMMARY

| Category | Existing Work |
|---|---|
| General Purpose Computers | Off-the-shelf [2][8][41][46] <br> Advanced multicore [42] <br> Manycore supercomputer [6] |
| GPU | [11][17][28][30][31][40] |
| Dedicated Hardware Solutions | FPGA [20][27] <br> Manycore Processor [29] <br> Distributed Array Processor [16] <br> Systolic Processor [34] <br> Coherent Processor [5] <br> TCAM / PIM [13] <br> Heterogeneous platform[32][33] <br> 3D LiM [35] |

The first category targets the optimization of sparse matrix multiplication on general purpose computer architectures. S. Toledo [41] enhanced sparse matrix multiplication on a superscalar RISC processor by improving instruction-level parallelism and reducing cache miss rate. A. Pinar et al. [2] proposed further optimization of data structures using reordering algorithms, to improve cache performance. E. Im et al. [8] developed the SPARSITY toolkit for the automatic optimization of sparse matrix multiplication. Y. Saad et al. [46] proposed PSPARSLIB, a collection of sparse matrix multiplication subroutines for multiprocessors. S. Williams et al. [42] examined and optimized sparse matrix multiplication across a broad spectrum of multicore architectures. Finally, Bowler et al. [6] optimized sparse matrix multiplication for a 512-core supercomputer.

Another direction is the implementation and optimization of sparse matrix multiplication using GPU. While this effort still relies on a conventional computational platform and focuses mainly on algorithm optimization, it enables significant speedup over sequential CPU or even multicore solutions [31]. Many of the GPU-based studies rely on G. Blelloch's [11] research into mapping of sparse data structures onto SIMD hardware. S. Sengupta et al. [40] developed segmented scan primitive for efficient sparse matrix multiplication on GPU. J. Bolz et al. [17] implemented a sparse matrix solver on GPU. M. Baskaran et al. [28] enhanced GPU sparse matrix multiplication by creating an optimized storage format. Bell et al. [30][31] develop methods to exploit common forms of matrix structure while offering alternatives to accommodate irregularity.

The third direction encompasses special purpose hardware solutions for sparse matrix multiplication. L. Zhuo [27] proposed an FPGA based design, which reportedly demonstrated a significant speedup over then-current general-purpose solutions (such as Itanium 2), especially for matrices with very irregular sparsity structures. Another FPGA based sparse matrix multiplication solution was introduced by J. Sun et al. [20]. Some specialty solutions relying on VLSI implementation have been suggested as well. M. Misra et al. [29] developed a parallel architecture comprising *nnz* processing elements (where *nnz* is the number of nonzero elements in a matrix), and implemented an efficient routing technique to resolve the communication bottleneck. J. Andersen et al. [16] suggested implementing sparse matrix multiplication on the Distributed Array Processor (DAP), a massively parallel SIMD architecture. O. Beaumont et al. [32][33] implemented matrix multiplication on a heterogeneous network.

A number of hardware solutions using content-addressable memory have also been proposed. O. Wing [34] suggested a systolic array architecture, comprising a number of processing elements connected in a ring. Each processing element has its own content-addressable memory, storing the nonzero elements of the sparse matrix. Matrix elements are extracted from the memory by content addressing. Sparse matrix-vector multiplication takes $O(nnz)$ cycles (where *nnz* is the number of nonzero elements in matrix). That work relies on an earlier study by R. Kieckhager et al. [37], who were probably the first to use a content-addressable memory in the context of sparse matrix multiplication. Q. Guo et al. [13] implemented a fixed point matrix multiplication on a TCAM based Processing-In-Memory (PIM) architecture. They use TCAM to match key-value pairs but rely on a microcontroller for multiplication. Recently, Q. Zhu et al. [35] suggested a 3-D Logic-In-Memory (LiM) architecture where DRAM dies are intertwined with logic dies in a 3D stack. Their architecture uses a logic-enhanced CAM to take advantage of its parallel matching capabilities.

Associative processors have also been considered in the context of matrix processing. C. Stormon [5] introduced the Coherent Processor, a massively parallel associative computer. Sparse matrix computations are mentioned among the Coherent Processor's applications although no details of the sparse matrix multiplication are provided. Stromon suggested using the Coordinate (COO) format of storing nonzero elements of sparse matrices along with their row and column indices, in con-trast other sparse formats such as Compressed Sparse Row (CSR) or ELLPACK (ELL) [18], which are more efficient for sequential processors or GPUs.

The key contribution of the present work is the efficient implementation of sparse matrix multiplication on a memory intensive associative processor (AP), verified by extensive AP simulation using a large collection of sparse matrices [43].

## 3 ASSOCIATIVE PROCESSOR

Associative Processor (AP) combines data storage and data processing, and functions as data memory and massively parallel SIMD accelerator at the same time. The architecture of AP and principles of associative computing is presented in [24].

Three out of four sparse by sparse matrix multiplication algorithms presented in this paper require a CPU operating in conjunction with the AP. The top-level architecture of the AP with an external CPU is shown in Fig. 1. The CPU is a baseline microprocessor, similar to that of [13], with a simple 4 stage pipeline and a typical instruction set including arithmetic/logic, memory access and control instructions. A single-precision floating point addition and multiplication in the CPU is assumed to be performed in a single pipeline stage.

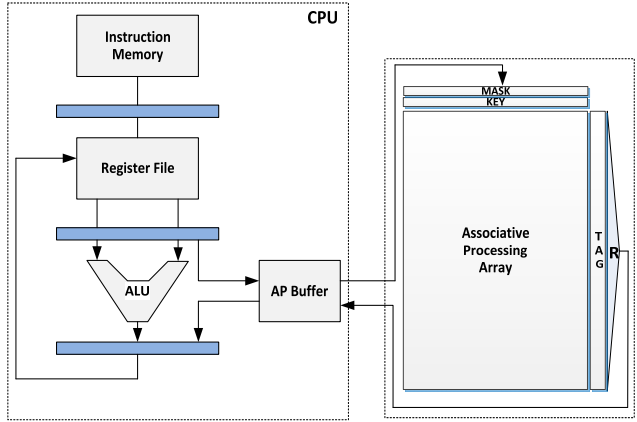

Fig. 1. AP with an external CPU

## 4 SPARSE MATRIX MULTIPLICATION ON AP

In this section we detail the sparse matrix multiplication algorithm and its four implementations on the AP.

Fig. 2 illustrates the multiplication of sparse matrix A by sparse matrix B. In this example, row $j$ of matrix A has three nonzero elements in columns $\{i_1, i_2, i_3\}$. Rows $\{i_1, i_2, i_3\}$ of matrix B have nonzero elements in columns $\{k_1, k_3, k_5\}$, $\{k_2, k_4\}$ and $\{k_1, k_2, k_5\}$, respectively.

Fig. 3 shows the associative processing array and reduction tree [25] mapping. We assume that both input matrices are stored in the AP in the COO format, where nonzero elements are entered consecutively, with the row and column indexes stored alongside the matrix element.

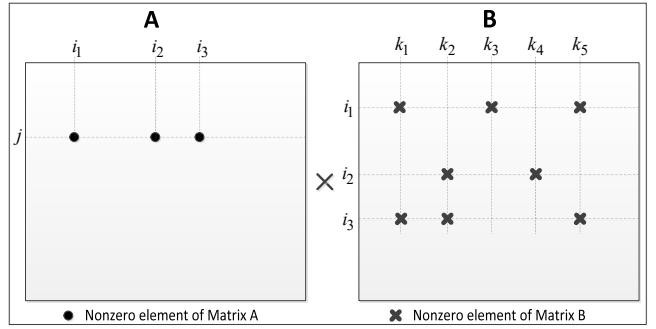

Fig. 2. Sparse Matrix Multiplication - Illustration

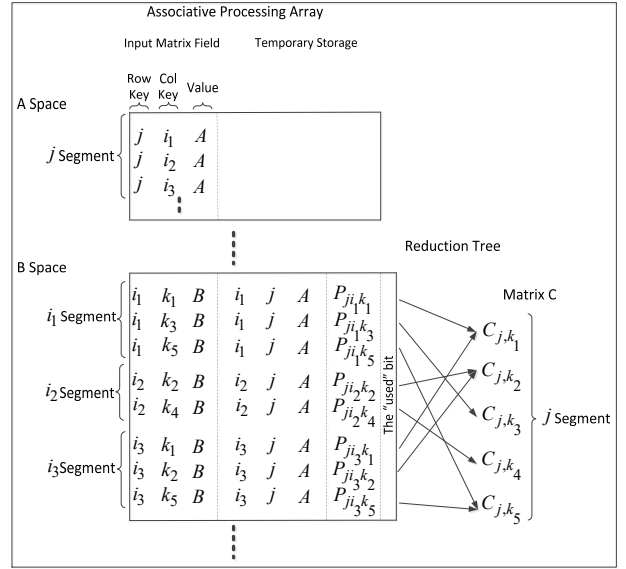

Fig. 3. AP Memory and Reduction Map

Fig. 4 presents the pseudo code of the fully associative sparse matrix multiplication (algorithm "AP"). It includes two internal loops nested within an external one. The external loop goes over the nonzero rows of matrix A. The first internal loop goes over the nonzero elements in each nonzero row of matrix A and takes three steps. At step 1, a nonzero element of row $j$ and its column index $i$ are read from the associative memory (associative processing array). At step 2, its column index $i$ is compared against the row index field of the entire matrix B. This step is done in parallel for all nonzero elements of matrix B, using the AP compare command. All matching nonzero elements of matrix B ($k_1$, $k_3$ and $k_5$ for row $i_1$ etc. in Fig. 2 and Fig. 3) are tagged. At step 3, the nonzero element of matrix A is written simultaneously into all tagged rows, alongside the tagged elements of matrix B (segments $i_1$, $i_2$ and $i_3$ of Fig. 3).

The first internal loop is repeated while there are nonzero elements in row $j$ of matrix A. Upon completion, all nonzero pairs of matrices A and B required to calculate the row $j$ of the product matrix C are aligned (stored in the same associative processing unit) in the associative processing array.

Next step 4 is the associative multiplication of A,B pairs, performed in parallel for all pairs. For instance, the index of the first product in Fig. 3 is $j, i_1, k_1$.

The second loop sums up the products (the singletons). It contains steps 5 through 8. At step 5, a singleton product is read from the associative processing array (beginning with the first one). At step 6, its B column index $k$ (unless it is marked "used") is compared against the B column index of all singleton products, and all singletons with B column index $k$ are tagged. At step 7, the tagged rows are marked "used" by a write command. Those tagged rows hold the singleton products that need to be accumulated to form element $C_{j,k}$. Step 8 is the reduction. The reduction tree is pipelined hence the loop may end without waiting for the reduction tree to complete. The loop is repeated while there are unprocessed (that is, not marked "used") B column indices.

```
Init     {
             Matrix A → A space;
             Matrix B → B space;
}

Main {
    While (!end of A) { //serially over all nz rows of A
        While (!end of row j) { //serially, over all nz elements in jth row of A
            1.  Read_next {i, A_{j,i}}
            2.  Tag all B_{i,k}   //in parallel, single step, for all k
            3.  Write A_{j,i}    //in parallel, single step, into all tagged rows
        }
        4.   P_{j,i,k} = ASSOCIATIVE_MULT(A_{j,i}, B_{i,k})  //forall aligned pairs

        While (∃k not used)  { //serially over all values k
            5.  Read_next {k, P_{j,i,k}}  //find next not used k value
            6.  Tag all P_{j,i,k}   //parallel forall P_{j,*,k} with same k, single step
            7.  Mark "used" //parallel forall tagged rows, single step
            8.  C_{j,k} = ASSOCIATIVE_REDUCE_SUM(P_{j,i,k})
        }
    }
}
```
Fig. 4. AP algorithm for fully associative sparse matrix multiplication

In certain sparse matrices, most rows and columns contain very few nonzero elements. In such cases, parallel reduction (step 8 in Fig. 4) may be less efficient because a very few singleton products are accumulated in each iteration. Consequently, the reduction may better be carried out word-serially, by the external CPU (Fig. 1). That algorithm, "AP+ACC," is shown in Fig. 5. Steps 1 through 6 are identical to those of "AP". The 8th step is a nested loop that goes over all the singleton products tagged at step 6. Each $P_{j,i,k}$ singleton is read and accumulated by an external CPU. We assume a pipelined operation so that steps 8a and 8b in Fig. 5 are performed in parallel; once the pipeline is filled, each pass of the loop takes a single cycle.

```
Same code as in Fig. 6, except:
    8.  Forall tagged rows // serially
        a.  CPU read P_{j,i,k}
        b.  C_{j,k}=CPU_ACC (C_{j,k}, P_{j,i,k})
```
Fig. 5. "AP+ACC" algorithm, using serial accumulation

Similarly, a parallel associative multiplication (step 4 in Fig. 4) may be inefficient when the average number of nonzero elements per matrix row is small. In such case, the multiplication of matrix elements may be best performed word-serially by an external CPU. Fig. 6 presents the pseudo code of this "AP+MULT" algorithm. Steps 1, 2 and 5 through 8 are identical to those of "AP". The 3rd step is a nested loop that goes over all the elements of matrix B with the row index matching the column index $i$ of the nonzero element $A_{j,i}$. Each $B_{i,k}$ element is multiplied by $A_{j,i}$ at the external CPU and is written back to the corresponding row of the associative processing array. We assume a pipelined operation so that steps 3b and 3c in Fig. 6 are performed in parallel; once the pipeline is filled, each pass of the loop takes 2 cycles.

```
Same as Fig. 6, except:

    3.  Forall B_{i,k}  // serially
        a.  Read_next B_{i,k};      // single step
        b.  P_{i,k}=CPU_MULT (A_{j,i}, B_{i,k})
        c.  Write P_{i,k} alongside B_{i,k};  // single step
line 4 is deleted
```
Fig. 6. "AP+MULT" algorithm using serial multiplication

Both algorithms "AP+MULT" and "AP+ACC" are combined into "AP+MULT+ACC" in Fig. 7. This algorithm is efficient for smaller matrices with a lower average number of nonzero elements per row (for example, diagonal matrices).

```
Init     {
             Matrix A → A space;
             Matrix B → B space;
}

Main {
    While (!end of A) { //serially over all nz rows of A
        While (!end of row j) { //serially, over all nz elements in jth row of A
            1.  Read_next {i, A_{j,i}}
            2.  Tag all B_{i,k}   //in parallel, single step, for all k
            3.  Forall B_{i,k} // serially
                a.  Read_next B_{i,k};      // single step
                b.  P_{i,k}=CPU_MULT (A_{j,i}, B_{i,k})
                c.  Write P_{i,k} alongside B_{i,k}; // single step
        }
        4.   P_{j,i,k} = ASSOCIATIVE_MULT(A_{j,i}, B_{i,k})  //forall aligned pairs

        While (∃k not used)  { //serially over all values k
            5.  Read_next {k, P_{j,i,k}}  //find next not used k value
            6.  Tag all P_{j,i,k}   //parallel forall P_{j,*,k} with same k, single step
            7.  Mark "used" //parallel forall tagged rows, single step
            8.  Forall tagged rows // serially
                a.  CPU read P_{j,i,k}
                b.  C_{j,k}=CPU_ACC (C_{j,k}, P_{j,i,k})
        }
    }
}
```
Fig. 7. "AP+MULT+ACC" algorithm using serial multiplication and accumulation

## 5 SIMULATIONS OF SPARSE BY SPARSE MATRIX MULTIPLICATION ON AP

The AP simulator [25] is used to quantify the efficiency of the four algorithms of Section 3. The experimental set-up, matrix statistics and simulation results are described in this section.

## 5.1 Experimental Setup

To simulate sparse matrix multiplication, we use up to 700 square matrices with the number of nonzero elements spanning from hundred thousand to eight million, randomly selected from the collection of sparse matrices from the University of Florida [43].

We simulate the sparse matrix multiplication using the AP simulator [25]. As shown in Fig. 3, each pair of matrix elements and the resulting singleton product are processed by a single AP processing unit. Simulations are performed on Intel® Core™ i7-3820 processor with 32GB RAM, and simulation times for the 100K−8M nonzero element matrices range between few minutes and few tens of hours.

For power simulation, we follow the methodology described in [25]. During AP execution, we record and count all baseline operations: match, mismatch, write, miswrite, reduction. Using power models of each baseline operation, detailed in [25], we estimate the total energy consumed during execution of each case.

## 5.2 Matrix Statistics

AP performance depends on the data wordlength rather than on data set size. Data wordlength is the maximum number of bits in fixed point representation of matrix elements. If matrix elements are presented in a floating point format, the wordlength is 32 bit (IEEE754 single precision). Data set size in SpMM typically equals the number of nonzero elements in the sparse matrix.

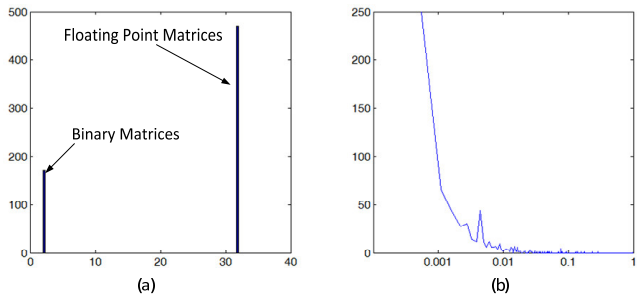

Fig. 8. (a) Wordlength histogram, (b) Histogram of the average number of nonzero elements per row, relatively to the matrix dimension

Fig. 8(a) presents the matrix element wordlength histogram. The first peak represents the binary matrices (two bits stand for a value bit and a sign). The second peak encapsulates matrices with floating point data elements. In this work, all multiplications are carried out as either binary (Boolean) or floating point operations.

There are several applications that use sparse binary matrices. According to [43], these applications may include recommender systems, undirected graph sequencing, certain optimization problems, duplicate structural problems, random un-weighted graph processing and computational fluid dynamics problems. To emphasize the efficiency of the "AP" algorithm, we employ parallel Boolean multiplication in the binary matrices: it takes only eight cycles, regardless of the number of nonzero elements in a row.

As we show in Section 5.3 below, the performance of the fully associative "AP" algorithm is strongly affected by the average number of nonzero elements per row. The distribution of the average number of nonzero elements per row relative to the matrix dimension is shown in Fig. 8(b).

In "AP" and "AP+ACC" algorithms, we calculate the singleton products by associatively multiplying the matrix elements. Consider a matrix containing a limited number of unique elements, known in advance. In such case, the products of all unique elements can be pre-calculated, and a "vocabulary" containing all pairs of the unique elements and their products can be created. Instead of multiplication, the AP would then match the pairs of the unique elements and substitute the pre-calculated product in the result field. For $n$ unique elements in a matrix, such vocabulary-based multiplication would take $2n^2$ cycles. Hence, if $2n^2$ is shorter than the associative multiplication time (in cycles), the "AP" and "AP+ACC" algorithms can be sped up by replacing associative multiplication by vocabulary-based one.

Fig. 9 shows the distribution of the $2n^2$ figure. The first peak corresponds to binary matrices and should therefore be excluded from the analysis. Matrices with $2n^2$ value left of the 8,800 mark (the associative floating point multiplication cycle count) on the horizontal axis form a group for which vocabulary multiplication is preferred. For the rest of the matrices, the number of the unique elements $n$ is too large for the vocabulary multiplication to be time-efficient. The percentage of matrices with the number of unique elements in the left field (excluding binary matrices) is around 15%.

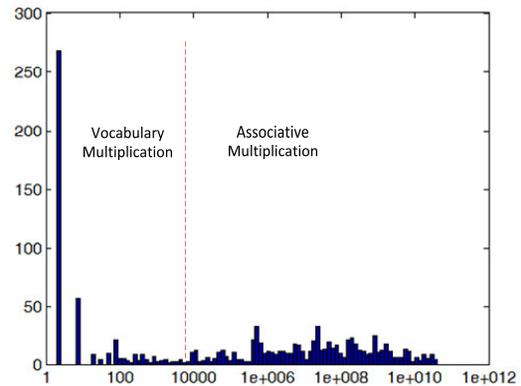

Fig. 9. $2n^2$ histogram, showing number of matrices having $n$ unique elements

We do not implement the vocabulary multiplication in our simulations, but find it worth noticing as an additional potential benefit of associative processing as compared to a conventional (GPU or multicore based) matrix multiplication.

## 5.3 Simulation Results

Fig. 10 presents the sparse by sparse matrix multiplication execution time of the four associative algorithms of Section 4 for the matrices with floating point elements (a) and with binary elements (b).

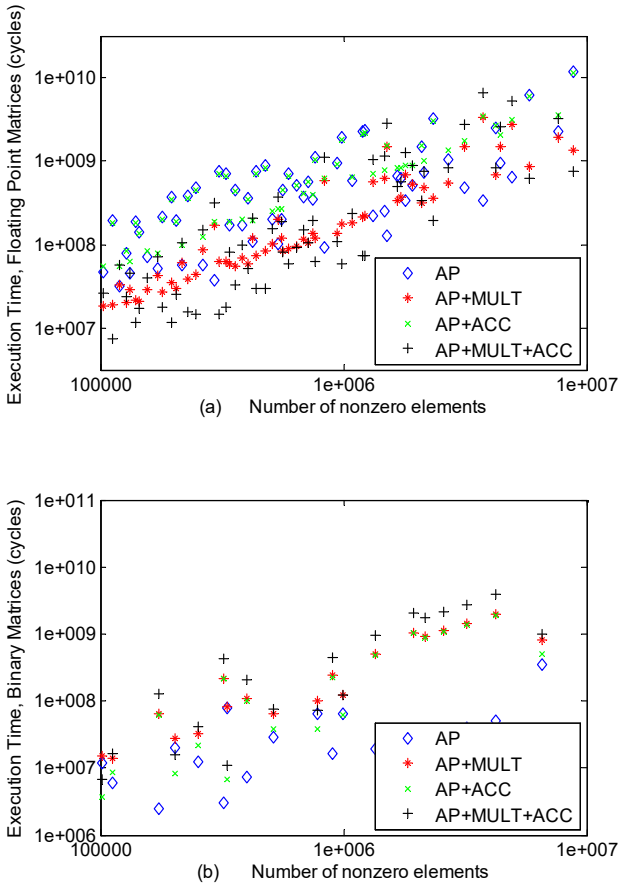

Fig. 10. Execution time vs. number of nonzero elements: (a) Floating point matrices; (b) Binary matrices

The reason for the spread in execution times (per each number of nonzero elements) in each individual algorithm is the sensitivity of the associative implementation to the average number of nonzero elements per row. For two matrices with a similar number of nonzero elements, the difference of two orders of magnitude in the average number of nonzero elements per row cause a similar difference in the execution time. For example, the "Williams/webbase-1M" matrix has 3,105,536 nonzero elements and an average of 3.1 nonzero elements per row. The "ND/nd3k" matrix however has 3,279,690 nonzero elements but an average of 364.4 nonzero elements per row. The multiplication of each of those two matrices by itself using the "AP" algorithm takes 8.7 and 0.17 billion cycles respectively, a difference of almost two orders of magnitude.

This sensitivity of performance to the average number of nonzero elements per row is shared, although possibly to a lesser extent, by conventional SpMV and SpMM implementations (on GPU or multicore) [19][44].

The difference in execution times of the "AP" algorithm with respect to binary vs. floating point matrices is a result of the difference in Boolean vs. associative multiplication times.

For smaller matrices (having less than one million nonzero elements), the "AP+MULT" and "AP+MULT+ACC" algorithms seem to provide the best performance in most cases, with the exception of binary matrices. For binary matrices, the picture is mixed. Even for the smallest matrices, the "AP" normally outperforms the hybrid algorithms, due to time-efficient Boolean multiplication.

As the number of nonzero elements approaches one million, the performance of the "AP" algorithm gradually improves. For matrices of several millions of nonzero elements, "AP" tends to outperform the hybrid algorithms.

This observation is also reflected in Fig. 11 and Fig. 12 that show the breakdown of the execution time into three major processing steps (pair matching, multiplication and accumulation) for all four algorithms. For example, for the Bourchtein/atmosmodj matrix [43] (with $nnz = 8,814,880$ nonzero floating point elements and average of $nnz\_per\_row = 7$ nonzero elements per row), the pair matching, multiplication and accumulation take approximately 0.002, 10.966 and 0.973 billion cycles respectively.

The analytical model for those three processing steps is as follows. The matching time is $2 \cdot nnz$ cycles. The average multiplication time is $nnzrow \cdot T_{mult}$ where $nnzrow = nnz/nnz\_per\_row$ is the average number of nonzero rows and $T_{mult}$ is the multiplication time (which for floating point matrix elements is 8,800 cycles). The average accumulation time is $nnz \cdot m + nnzrow \cdot T_{red}$ where $m$ is the data wordlength (32 bit for single precision floating point elements), and $T_{red}$ is the reduction time (which for floating point matrix elements is 600 cycles). The average execution time of the 'AP' algorithm is the sum of the above three components.

Figures Fig. 11 and Fig. 12 also show the execution time of matrix multiplication performed entirely on the baseline CPU of Fig. 1 (without AP acceleration). The execution times in Fig. 11(a) and (b) are averaged among all floating point and all binary matrices respectively. The average execution times for ten largest floating point and binary matrices are presented in in Fig. 12(a) and (b) respectively. In all cases, the pair matching step takes considerably longer on CPU than on the AP, where it is sped up by parallel execution. While for binary matrices, the "AP" algorithm exhibits the lowest delay on average, for the floating point matrices the picture is more complicated. The lowest average execution time over all floating point matrices is achieved by the "AP+MULT" algorithm. However, for the ten largest floating point matrices, the "AP" algorithm shows the best performance, as also indicated in Fig. 10. This has to do with the average number of nonzero elements per row which tends to be higher for larger matrices.

The performance of the "AP" algorithm, along with performances of the hybrid algorithms, for floating point and binary matrices, as functions of the number of nonzero elements, are presented in Fig. 13(a) and (b), respectively. For comparison, Fig. 13 also shows the SpMM performance of Intel Xeon Phi and NVidia K20 [9].

The spread in "AP" performance is a function of the average number of nonzero elements per matrix row. The divergence between binary and floating point perfor-mance is a result of Boolean vs. associative multiplication time difference.

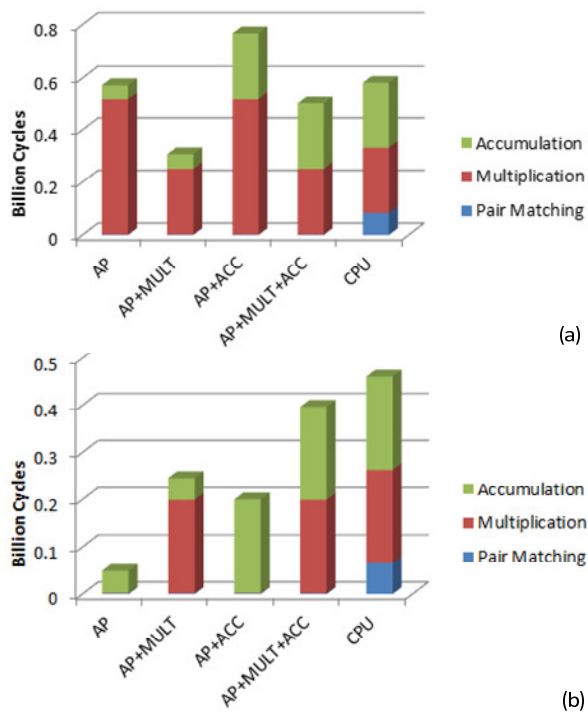

Fig. 11. Execution time breakdown: (a) All floating point matrices; (b) All binary matrices

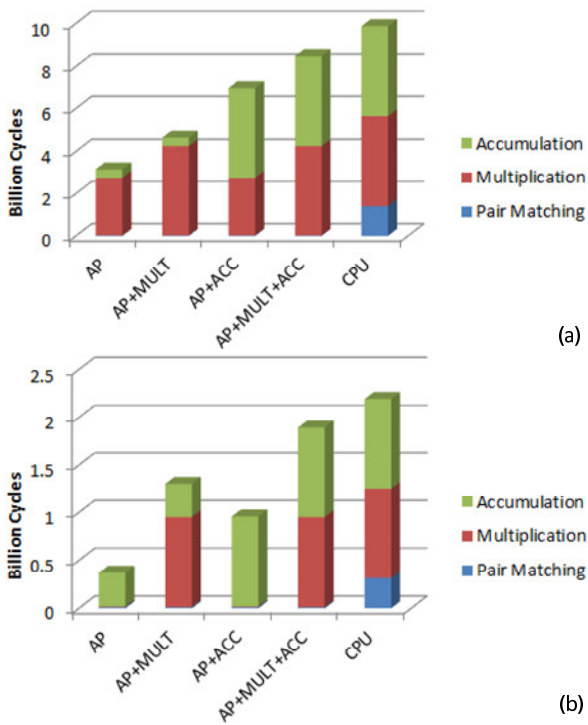

Fig. 12. Execution time breakdown: (a) Ten largest floating point matrices; (b) Ten largest binary matrices

The difference in performance of the "AP" sparse algorithm relative to the other two architectures shown in Fig. 13 is a result of a relative inefficiency of associative arithmetic when applied in parallel to small sets of numbers. As discussed below, the argument for using AP lies in its superior power efficiency, rather than in performance.

The performance of all four associative algorithms as functions of the average number of nonzero elements per row is presented in Fig. 14. The "AP" algorithm multiplies a row of the left matrix by the entire right matrix. Therefore the number of element by element multiplications performed in parallel is proportional to the number of nonzero elements in that row. Consequently the performance of "AP" algorithm exhibits almost linear dependency on the average number of nonzero elements per row, while the performance of the rest of the algorithms saturates, or even remains constant. Hence, if the average number of nonzero elements per row is small (which is consistently the case in University of Florida collection matrices), the effectiveness of the "AP" algorithm is limited. "AP" is least efficient for diagonal matrices, where there is only one multiplication per nonzero row. On the other end of the efficiency scale is dense matrix multiplication, where an associative multiplication is applied to $N^2$ matrix elements in parallel ($N$ is the matrix dimension) per each matrix row. For comparison, a 2000×2000 dense matrix multiplication performance is also shown in Fig. 14.

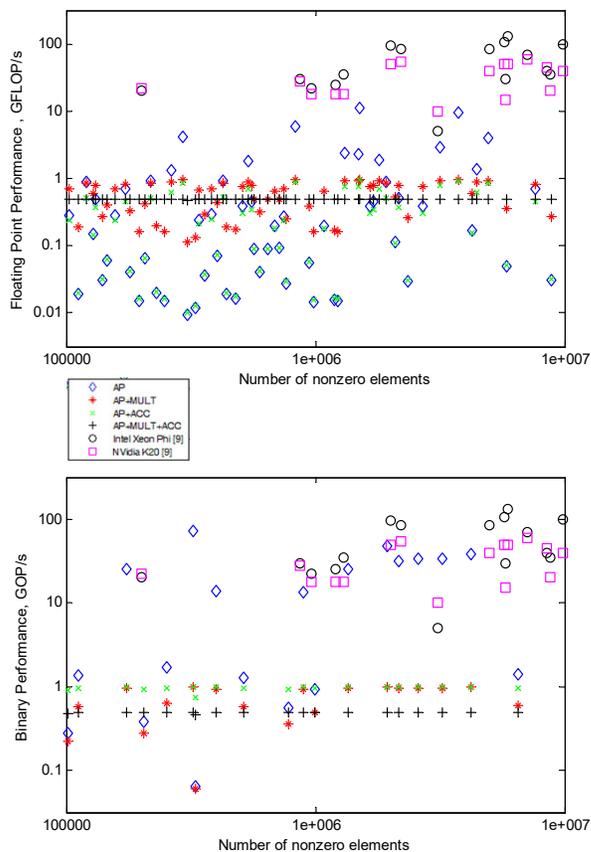

Fig. 13. Performance vs. number of nonzero elements: (a) Floating point matrices; (b) Binary matrices

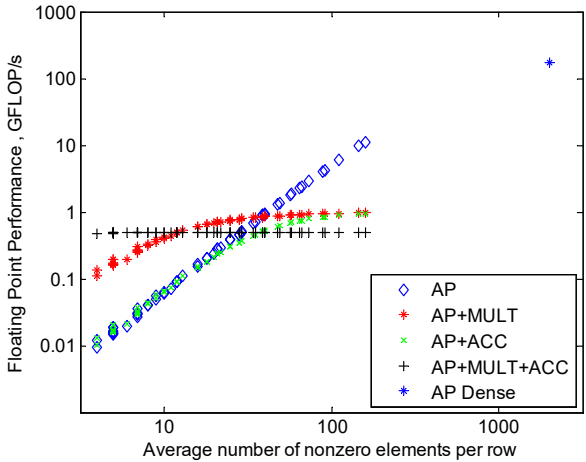

Fig. 14. Floating matrix SpMM performance vs. average number of nonzero elements per row

The simulated power consumption of the "AP" algorithm for floating point matrices as a function of the number of nonzero elements is presented in Fig. 15. For comparison, Fig. 15 also shows the estimated power consumption of NVidia K20 [36].

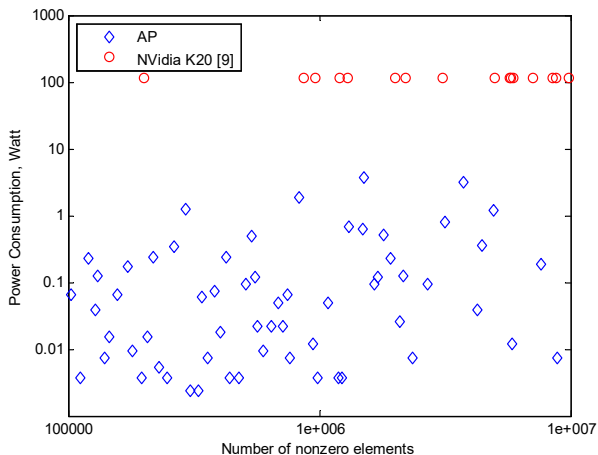

Fig. 15. Floating matrix SpMM power consumption vs. average number of nonzero elements per row

According to our simulations, the SpMM power efficiency of the AP is in the range of 5 to 10 GFLOP/s/W. The SpMM/SpMV power efficiency of the advanced contemporary GPUs such as NVidia's K20 and GTX660 is in the 0.1-0.5 GFLOP/s/W range [36]. A wide variety of multicore processors such as quad-core AMD Opteron 2214, quad-core Intel Xeon E5345, eight-core Sun UltraSparc T2+ T5140 and eight-SPE IBM QS20 Cell reportedly reach the SpMM power efficiency of up to 0.03 GFLOP/s/W [42]. The AP power efficiency advantage stems from in-memory computing (there are no data transfers between processing units and memory hierarchies) and from low-power design made possible by the very small size of each processing unit.

A noticeable limitation of the "AP" algorithm is the sequential processing of the matrix rows (the outer loop of Fig. 4). A parallelization of matrix row processing may significantly improve the performance of the "AP" algorithm. For example, diagonal matrices can easily be processed in a row-parallel manner, since there is only one nonzero singleton product per each matrix row. An optimization of the "AP" algorithm is the subject of our future work.

## 6 CONCLUSIONS

Sparse matrix multiplication is of great importance for many linear algebra applications, especially machine learning. The efficient implementation of sparse matrix multiplication becomes even more critical when applied to big data problems.

Associate Processor (AP) is essentially a large associative memory with massively-parallel processing capabilities. This paper investigates the merit of implementing sparse matrix multiplication on the AP.

We propose and compare four algorithms for the AP, from a fully associative computation to a hybrid of AP and CPU. To quantify the efficiency of the proposed algorithms, we simulate them using a large variety of sparse matrices.

We find that the fully associative "AP" algorithm has a computational complexity of $O(nnz)$ (where $nnz$ is the number of nonzero elements), and its efficiency grows with the number of nonzero elements per row. The "AP" algorithm multiplies in parallel a row vector of one matrix by the entire second matrix. As a result, the efficiency and performance of the "AP" algorithm also grows with the total number of nonzero elements.

We show that associative implementation can offer performance benefits when multiplying sparse matrices with a limited number of predefined unique elements. Lastly, we show that AP sparse by sparse matrix multiplication implementation is more power-efficient than conventional GPU or multicore based solutions. This is even more evident in the case of binary matrices, thanks to the bit-oriented nature of associative processing.

Associative implementation of sparse by sparse matrix multiplication may benefit from further optimization, such as parallelization of matrix row processing and possibly employing a combination of CSR and CSC instead of COO sparse matrix storage format.

## ACKNOWLEDGMENT

This research was partially funded by the Intel Collaborative Research Institute for Computational Intelligence and by Hasso-Plattner-Institut.

**Leonid Yavits** received his MSc in Electrical Engineering from the Technion. After graduating, he co-founded VisionTech where he co-designed a single chip MPEG2 codec. Following VisionTech's acquisition by Broadcom, he co-founded Horizon Semiconductors where he co-designed a Set Top Box on chip for cable and satellite TV.

Leonid is a PhD student in Electrical Engineering in the Technion. He co-authored a number of patents and research papers on SoC and ASIC. His research interests include Processing in Memory and 3D IC design.

**Amir Morad** received his BSc and MSc in Electrical Engineering from the Technion. Amir co-founded Vision-Tech, a major provider of ICs for set top boxes market. Following VisionTech's acquisition by Broadcom, Amir co-founded Horizon Semiconductors, where he co-designed SoCs for HD cable and satellite set top boxes.

Amir is a PhD student in Electrical Engineering in the Technion. He co-authored a number of patents and research papers on SoC and ASICs. His research interests include analytical modeling and optimization of many-core architectures.

**Ran Ginosar** received his BSc from the Technion and his PhD from Princeton University. After conducting research at AT&T Bell Laboratories, he joined the Technion where he is now professor at the Electrical Engineering department and a head of the VLSI Research Center.

Professor Ginosar has been a visiting Associate Professor with the University of Utah and co-initiated the Asynchronous Architecture Research Project at Intel (Oregon). He has co-founded a number of VLSI companies. Professor Ginosar has published numerous papers and patents on VLSI. His research interests include VLSI architecture, asynchronous logic and synchronization.